# ON THE HIGHER ORDER HAMILTON THEORY IN FIBERED SPACES

D. Krupka

**1. Introduction.** In this paper we will be considering a basic geometric problem, the extension problem of classical Hamilton-Cartan variational theory to higher jet prolongations on fibered manifolds.

Primary purpose of recent publications on this problem has been the extension of the first order Hamilton-Cartan theory as developed by Goldschmidt and Sternberg [7] and Dedecker [3] (see also [9]). A part of these publications are concerned with the search for "generalized canonical equations" [1], [2], [5], [11], [19], [20]. Most of them are based on a local generalization of the first order fundamental Cartan form, and turn up to follow the de Donder's approach to higher order variational problems [4]. Shadwick [20] has found, within the local theory, a satisfactory "regularity condition" approving the canonical equations which guarantees that these equations be equivalent with the Euler-Lagrange equations, and leads to the Hamilton-Jacobi equation for extremal fields in a similar way as in the first order theory. Some authors [6], [9], [11], [15], [18] have been concerned among other problems with the global existence of an appropriate "generalized Cartan form", that is, a global Lepagean equivalent of a given lagrangian in the sense of [13]. The author and Stepankova [14] considered a class of second order lagrangians, admitting a first order Lepagean equivalent.

In most of these publications, however, the concept of regularity, as well as the global aspects of the theory, have either remained indistinct or have been left quite aside. Although all of them have a certain stimulating meaning or contain new ideas, some of the results seem to be problematic and should therefore be re-examined.

The purpose of this paper is a brief discussion of the elements of the Hamilton-Cartan theory on the basis of Lepagean forms, introduced, in accordance with the ideas of Lepage [16], [17], by Krupka [10], [12], [13]. These forms reflect all the main properties of the Cartan form, and are naturally becoming the fundamental element for geometrization of the higher order calculus of variations in fibered spaces. Basic properties of the Lepagean forms are recalled in Section 3. In Section 4, we introduce the Hamilton form, associated with a Lepagean form; the construction is analogous to [14]. The Hamilton form is precisely the Euler-Lagrange form of an "extended" lagrangian, defined, in a canonical manner, by means of the initial lagrangian. Section 5 is concerned with the discussion of this form under a regularity assumption; the correspondence between extremals of the lagrangian and the extended lagrangian is studied. In Section 6 we discuss the Euler-Lagrange equations of the extended lagrangian, in the local and "regular" case; these equations rewritten is the "Legendre coordinates", coincide with the canonical equations given by de Donder [4]. Section 7 is devoted to some remarks on the theory of extremal fields; we extend the results obtained in [20].

**2. Notation.** Throughout this paper, $\pi: Y \to X$ is a fixed fibered manifold with oriented base $X$, and $\dim X = n$, $\dim Y = n + m$. The $r$-jet prolongation of $\pi$ is a fibered manifold denoted by $\pi_r : j^r Y \to X$, and $\pi_{r,s} : j^r Y \to j^s Y$, $0 \leq s \leq r$, are natural jet projections. A typical element of $j^r Y$ - the $r$-jet of a section $\gamma$ of $\pi$ at a point $x \in X$, is denoted by $j^r_x \gamma$. In accordance with this notation, we use the symbols $(\pi_r)_s : j^s (j^r Y) \to X$ for the $s$-jet prolongation of $\pi_r$, and $(\pi_r)_{s,k}$ for the corresponding natural jet projections. The $r$-jet prolongation $x \to j^r_x \gamma$ of a section $\gamma$ of $\pi$ is a section of $\pi_r$, denoted by $j^r \gamma$, and the $r$-jet prolongation of a $\pi$-projectable vector field $\xi$ on $Y$ is a vector field on $J^r Y$ denoted by $j^r \xi$.

If $(V, \psi)$ is a fiber chart on $Y$, with coordinates $\psi = (x^i, y^\sigma)$, $1 \leq i \leq n$, $1 \leq \sigma \leq m$, the chart on $j^r Y$ associated with the chart $(V, \psi)$ is denoted by $(V_r, \psi_r)$, $\psi_r = (x^i, y^\sigma, y^\sigma_{j_1}, \ldots, y^\sigma_{j_1 \ldots j_r})$; in these expressions, $1 \leq j_1 \leq \ldots \leq j_r \leq n$. The chart on $j^1(j^r Y)$ associated with $(V, \psi)$ is denoted by $((V_r)_1, (\psi_r)_1)$, $(\psi_r)_1 = (x^i, y^\sigma, y^\sigma_{j_1}, \ldots, y^\sigma_{j_1 \ldots j_r}, y^\sigma_{,p}, y^\sigma_{j_1,p}, \ldots, y^\sigma_{j_1 \ldots j_r,p})$; here $1 \leq p \leq n$, $1 \leq \sigma \leq m$, and again $1 \leq j_1 \leq \ldots \leq j_r \leq n$.

The following modules of forms are used in the text: the module of $p$-forms on $j^r Y$, denoted by $\Omega^p(j^r Y)$, the module of $\pi_r$-horizontal $p$-forms, denoted by $\Omega^p_X(j^r Y)$, and the module of $\pi_{r,s}$-horizontal $p$-forms, denoted by $\Omega^p_{j^s Y}(j^r Y)$. The letter $h: \Omega^p(j^r Y) \to \Omega^p_X(j^{r+1} Y)$ (resp. $h: \Omega^p(j^s(j^r Y)) \to \Omega^p(j^{s+1}(j^r Y))$) is used for the $\pi$-horizontalization (resp. $\pi_r$-horizontalization) mapping. If $\rho \in \Omega^p(j^r Y)$ then $h(\rho)$ may be defined as a unique element of $\Omega^p_X(j^{r+1} Y)$ such that $j^r \gamma^* \rho = j^{r+1} \gamma^* h(\rho)$ for each section $\gamma$ of $\pi$ (* denotes the pullback). A form $\rho \in \Omega^p(j^r Y)$ is called 1-contact if for each $\pi_r$-vertical vector field $\xi$ on $j^r Y$ the form $i_\xi \rho$ (the inner product of $\xi$ and $\rho$) is $\pi_r$-horizontal, i.e., belongs to $\Omega^{p-1}_X(j^r Y)$; $\rho$ is called $k$-contact, where $2 \leq k \leq p$, if $i_\xi \rho$ is $(k-1)$-contact. $q$-contact $p$-forms define a submodule of $\Omega^p_{j^{r-1}Y}(j^r Y)$, denoted by $\Omega^{p-q,p}(j^r Y)$. Each element $\rho \in \Omega^p_{j^{r-1}Y}(j^r Y)$ has a unique decomposition of the form $\rho = h(\rho) + p_1(\rho) + p_2(\rho) + \ldots + p_p(\rho)$, where $h(\rho) \in \Omega^p_X(j^r Y)$, and $p_i(\rho) \in \Omega^{p-i,i}(j^r Y)$; the order of contact of $\rho$ is said to be $\leq k$ if $p_{k+1}(\rho) = 0, \ldots, p_p(\rho) = 0$.

The field of real numbers is denoted by $\mathbf{R}$. If $f$ is a function on $j^r Y$ and $(V, \psi)$, $\psi = (x^i, y^\sigma)$, a fiber chart on $Y$, we have $h(df) = d_i f \, dx^i$ on $V_{r+1} \subset j^{r+1} Y$; the function $d_i f: V_{r+1} \to \mathbf{R}$ is the $i$-th formal derivative of $f$; analogously, if $f$ is a function on $j^s(j^r Y)$, we have $h(df) = d_i f \, dx^i$, which defines a function $d_i f: (V_r)_{s+1} \to \mathbf{R}$.

A form $\rho \in \Omega^p(j^s Y)$ is sometimes considered as an element of $\Omega^p(j^r Y)$, $\pi^*_{r,s} \rho$, where $r > s$; when there is no danger of confusion we denote both of these forms by the same letter, $\rho$.

Standard summation convention is used unless otherwise stated; on the other hand, non-standard summations over distinguished indices are always explicitly designated.

Notation and the notions introduced in this section, may be consulted with [10] or [11].

**3. Lagrangians and their Lepagean equivalents.** Recall that a *lagrangian of order* $r$ for $\pi$ is an element $\lambda \in \Omega^n_X(j^r Y)$. A form $\rho \in \Omega^n(j^s Y)$ is called a *Lepagean equivalent* of a lagrangian of order $r$, if (1) $h(\rho) = \lambda$, and (2) $p_1(d\rho)$ is a $\pi_{s,0}$-horizontal form. The *action function* of $\lambda$ over a piece $\Omega \subset X$ is the real-valued function $\gamma \to \lambda_\Omega(\gamma) = \int_\Omega j^r \gamma^* \lambda$ defined on the set of sections of $\pi$ over $\Omega$.

The following theorem provides a criterion for a form of a special kind to be a Lepagean equivalent of a lagrangian. We formulate it here with a different summation convention than in [11]. Let us denote

(3.1) $$\begin{aligned} \omega_0 &= dx^1 \wedge \ldots \wedge dx^n, \\ \omega_i &= (-1)^{i-1} dx^1 \wedge \ldots \wedge dx^{i-1} \wedge dx^{i+1} \wedge \ldots \wedge dx^n, \\ \omega^\sigma_{j_1 \ldots j_k} &= dy^\sigma_{j_1 \ldots j_k} - y^\sigma_{j_1 \ldots j_k l} dx^l. \end{aligned}$$

**Theorem 1.** *Let* $\rho \in \Omega^n_{j^{r-1}Y}(j^{2r-1} Y)$ *be a form whose order of contact is* $\leq 1$. $\rho$ *is a Lepagean equivalent of a lagrangian* $\lambda \in \Omega^n_X(j^r Y)$ *if and only if for any fiber chart* $(V, \psi)$, $\psi = (x^i, y^\sigma)$, *on* $Y$

(3.2) $$\rho = L\omega_0 + \sum_{k=0}^{r-1} \sum_{j_1 \leq \ldots \leq j_k} f^{i,j_1 \ldots j_k}_\sigma \omega^\sigma_{j_1 \ldots j_k} \wedge \omega_i,$$

*where* $L$ *is defined by the chart expression*

(3.3) $$\lambda = L\omega_0,$$

*and* $f^{i,j_1 \ldots j_k}_\sigma$ *are functions on* $V_{2r-1}$, *defined by*

(3.4) $$\frac{1}{N(j_1 \ldots j_{r-1})} f^{i_r, j_1 \ldots j_{r-1}}_\sigma = \frac{1}{N(j_1 \ldots j_r)} \frac{\partial L}{\partial y^\sigma_{j_1 \ldots j_r}} + g^{j_r, j_1 \ldots j_{r-1}}_\sigma,$$

$$\frac{1}{N(j_1\ldots j_{k-1})} f_\sigma^{i_k, j_1\ldots j_{k-1}}$$

$$= \frac{1}{N(j_1\ldots j_k)}\left(\frac{\partial L}{\partial y^\sigma_{j_1\ldots j_k}} - d_i f_\sigma^{i,j_1\ldots j_k}\right) + g_\sigma^{j_k, j_1\ldots j_{k-1}}, \quad 2 \le k \le r-1,$$

$$f_\sigma^{j_1} = \frac{\partial L}{\partial y^\sigma_{j_1}} - d_i f_\sigma^{i,j_1},$$

*where*

(3.5)     $g_\sigma^{(j_k, j_1\ldots j_{k-1})} = 0, \quad 2 \le k \le r.$

**Proof.** Theorem 1 follows from the chart expressions of the form $\rho$ and $d\rho$, and from the definition of the Lepagean equivalent.

With the notation of Theorem 1, set

(3.6)
$$P_\sigma^{j_1\ldots j_r} = \frac{1}{N(j_1\ldots j_r)} \frac{\partial L}{\partial y^\sigma_{j_1\ldots j_r}},$$

$$P_\sigma^{j_1\ldots j_k} = \frac{1}{N(j_1\ldots j_k)} \frac{\partial L}{\partial y^\sigma_{j_1\ldots j_k}} - d_i P_\sigma^{ij_1\ldots j_k}, \quad 1 \le k \le r-1,$$

and

(3.7)
$$\Theta_{\lambda,V} = L\omega_0 + \sum_{k=0}^{r-1} \sum_{j_1 \le \ldots \le j_k} N(j_1\ldots j_k) P_\sigma^{ij_1\ldots j_k} \omega^\sigma_{j_1\ldots j_k} \wedge \omega_i$$

$$= L\omega_0 + \sum_{k=0}^{r-1} P_\sigma^{ij_1\ldots j_k} \omega^\sigma_{j_1\ldots j_k} \wedge \omega_i$$

$\Theta_{\lambda,V}$ is a Lepagean equivalent of $\lambda$ restricted to $V_r$; we call it the *local Poincare-Cartan equivalent* of $\lambda$.

The problem of existence of a Lepagean equivalent of a lagrangian is solved affirmatively in the following theorems.

**Theorem 2.** *Each lagrangian of order $r$ has a Lepagean equivalent.*

**Proof.** We take a system $(V_\iota, \psi_\iota)$, $\iota \in I$, of fiber charts on $Y$ such that the open sets $V_\iota$ define a locally finite covering of $X$, and a partition of unity $(\chi_\iota), \iota \in I$, subordinate to this covering. Let $\lambda \in \Omega^n_X(j^r Y)$ be a lagrangian. For any $\iota$, the local Poincare-Cartan equivalent $\Theta_{\chi_\iota \lambda, V_\iota}$, may then be regarded as a globally well-defined form. The proof consists in checking that the form

(3.8)     $\rho = \sum_\iota \Theta_{\chi_\iota \lambda, V_\iota}$

is a Lepagean equivalent of $\lambda$.

**Theorem 3.** *Let* $\lambda \in \Omega_X^n(j^r Y)$ *be a lagrangian. There exists a Lepagean equivalent* $\rho$ *of* $\lambda$ *such that for each fiber chart* $(V, \psi)$, $\psi = (x^i, x^\sigma)$, *on* $Y$

(3.9) $\quad \rho = \Theta_{\lambda,V} + \nu_V,$

*where* $\Theta_{\lambda,V}$ *is defined by* (3.7), *and*

$$\nu_V = \left( Q_\sigma^j \omega^\sigma + \sum_{k=1}^{r-2} \sum_{j_1 \leq \ldots \leq j_{k-1}} N(j_1 \ldots j_k) Q_\sigma^{j,j_1\ldots j_k} \omega^\sigma_{j_1\ldots j_k} \right) \wedge \omega_j,$$

(3.10)
$$Q_\sigma^{j_{r-1},j_1\ldots j_{r-2}} = g_\sigma^{j_{r-1},j_1\ldots j_{r-2}},$$

$$Q_\sigma^{j_k,j_1\ldots j_{k-1}} = g_\sigma^{j_k,j_1\ldots j_{k-1}} + \sum_{s=1}^{r-k-1}(-1)^s d_{i_1}\ldots d_{i_s} g_\sigma^{i_s,i_1\ldots i_{s-1}j_1\ldots j_k}, \quad 2 \leq k \leq r-2,$$

$$Q_\sigma^j = \sum_{s=1}^{r-2}(-1)^s d_{i_1}\ldots d_{i_s} g_\sigma^{i_s,i_1\ldots i_{s-1}j},$$

*where* $g_\nu^{p_k,p_1\ldots p_{k-1}}$ *are functions of the coordinates* $x^i$, $y^\sigma$, $y^\sigma_{j_1}$, ..., $y^\sigma_{j_1\ldots j_{2r-1-k}}$ *only such that*

(3.11) $\quad g_\sigma^{(j_k,j_1\ldots j_{k-1})} = 0.$

**Proof.** To prove this theorem, one should specify the components $f_\sigma^{i,j_1\ldots j_k}$ of $\rho$ (3.2), where $\rho$ is constructed by (3.8). It is essential to notice that the form $\nu_V$ in (3.9) is a Lepagean equivalent of the zero lagrangian.

The meaning of the theory of Lepagean forms for the calculus of variations consists, roughly speaking, in the geometrization of the theory. Let $\partial_\zeta$ denote the Lie derivative with respect to a vector field $\zeta$. For any compact piece $\Omega \subset X$, $\pi$-projectable vector field $\xi$ on $Y$, and any section $\gamma$ of $\pi$ over $\Omega$,

(3.12) $\quad \int_\Omega j^r \gamma^* \partial_{j^r \xi} \lambda = \int_\Omega j^{2r-1} \gamma^* i_{j^{2r-1}\xi} d\rho + \int_{\partial \Omega} j^{2r-1} \gamma^* i_{j^{2r-1}\xi} \rho,$

which is the *first variation formula;* $\gamma$ is an extremal of $\lambda$ if and only if

(3.13) $\quad j^{2r-1} \gamma^* i_{j^{2r-1}\xi} d\rho = 0$

for each $\xi$. The definition of a Lepagean equivalent of $\lambda$ immediately implies that (3.13) is equivalent, locally, with the system of Euler-Lagrange equations for $\gamma$.

**4. The Hamilton form.** Let $\lambda \in \Omega_X^n(j^r Y)$ be a lagrangian, $\rho$ a Lepagean equivalent of $\lambda$; assume that $\rho \in \Omega^n(j^{2r-1}Y)$. There is one and only one form $H_\rho$ on $j^1(j^{2r-1}Y)$ such that

(4.1) $\quad i_{j^1\xi} H_\rho = h(i_\xi d\rho)$

for all $\pi_{2r-1}$-vertical vector fields $\xi$ on $j^{2r-1}Y$. We call this form the *Hamilton form*

of $\rho$. A section $\delta$ of $\pi_{2r-1}$ is called a *Hamilton extremal* of $\rho$ if

(4.2)    $\delta^* i_\xi d\rho = 0$.

Obviously, $\delta$ is a Hamilton extremal if and only if $H_\rho \circ j^1\delta = 0$.

If $\gamma$ is an extremal of $\lambda$ then $\delta = j^{2r-1}\gamma$ is a Hamilton extremal of $\rho$.

**Theorem 4.** *The Hamilton form $H_\rho$ coincides with the Euler-Lagrange form of the first order lagrangian $\lambda \in \Omega_X^n(j^1(j^{2r-1}Y))$ defined by*

(4.3)    $\lambda = h(\rho)$.

**Proof.** Since the Lie derivative $\partial_{j^1\xi}$ commutes with the $\pi_{2r-1}$-horizontalization $h$, we have for every $\pi_{2r-1}$-vertical vector field $\xi$ on $j^{2r-1}Y$

(4.4)    $\partial_{j^1\varphi} h(\rho) = h(\partial_\varphi \rho) = h(i_\varphi d\rho) + h(di_\varphi \rho) = i_{j^1\varphi} H_\rho + h(di_\varphi \rho)$.

Since $\rho$ may be regarded as a Lepagean equivalent of the first order lagrangian $\lambda = h(\rho)$, (4.4) becomes the infinitesimal first variation formula for $\lambda$, and $H_\rho$ must be the Euler-Lagrange form of $\lambda$.

We can use Theorem 4 to derive the chart expression for $H_\rho$. If $\rho$ is expressed by (3.2), we have by definition

(4.5)    $\lambda = L\omega_0$,

where

(4.6)    $L = L + \sum_{k=0}^{r-1} \sum_{j_1 \leq \ldots \leq j_k} f_\sigma^{i,j_1\ldots j_k}(y^\sigma_{j_1\ldots j_k,i} - y^\sigma_{j_1\ldots j_k,i})$.

Then

(4.7)    $H_\rho = \sum_{s=0}^{2r-1} \sum_{p_1 \leq \ldots \leq p_s} H_\nu^{p_1\ldots p_s} dy^\nu_{p_1\ldots p_s} \wedge \omega_0$,

where

(4.8)    $H_\nu^{p_1\ldots p_s} = \dfrac{\partial L}{\partial y^\nu_{p_1\ldots p_s}} - d_q \dfrac{\partial L}{\partial y^\nu_{p_1\ldots p_s,q}}$.

**5. Regular lagrangians.** Consider a lagrangian $\lambda \in \Omega_X^n(j^r Y)$. A point $j_x^{2r-1}\gamma \in j^{2r-1}Y$ is said to be *regular* (relative to $\lambda$) if there exists a fiber chart $(V,\psi)$, $\psi = (x^i, y^\sigma)$, on $Y$ such that $\gamma(x) \in V$, and the functions $x^i$, $y^\sigma$, $y^\sigma_{j_1}$, ..., $y^\sigma_{j_1\ldots j_{r-1}}$, $P_\sigma^{j_1}$, ... $P_\sigma^{j_1\ldots j_r}$ (3.6), where $j_1 \leq j_2 \leq \ldots \leq j_r$, form a part of a chart $(W,\Psi)$ on $j^{2r-1}Y$ about $j_x^{2r-1}\gamma$. $\lambda$ is called *regular* if every point $j_x^{2r-1}\gamma \in j^{2r-1}$ is regular.

We shall now study the system (4.2) locally, taking $\Theta_{\lambda,V}$ instead of $\rho$, under the hypothesis that $\lambda$ is regular.

**Theorem 5.** *Let $\lambda \in \Omega_X^n(j^r Y)$ be a regular lagrangian, $\gamma$ a section of $\pi$. Suppose that to each point $x \in X$ from the domain of definition of $\gamma$ there exists a fiber chart $(V, \psi)$ on $Y$ and a section $\delta$ of $\pi_{2r-1}$ over $U = \pi(V)$ such that $\gamma = \pi_{2r-1} \circ \delta$ on $U$, and*

$$(5.1) \qquad \delta^* i_\varphi d\Theta_{\lambda,V} = 0$$

*for each $\pi_{2r-1}$-vertical vector field $\xi$ on $j^{2r-1}Y$. Then $\gamma$ is an extremal of $\lambda$.*

**Proof.** Let $H_{\lambda,V}$ be the Hamilton form of $\Theta_{\lambda,V}$. By (4.1),

$$(5.2) \qquad j^1 \delta^* i_{j^1\varphi} H_{\lambda,V} = \delta^* i_\varphi d\Theta_{\lambda,V}$$

for each section $\delta$ of $\pi_{2r-1}$ over $U = \pi(V)$, and each $\pi_{2r-1}$-vertical vector field $\xi$ on $V_{2r-1}$. If a section $\delta$ satisfies (5.1), we have

$$(5.3) \qquad H_\nu^{p_1 \ldots p_s} \circ j^1 \delta = 0, \quad 0 \leq s \leq 2r-1,$$

where the expressions on the left are determined by (4.9), (4.10). The equations (5.3) are linear in $y_{j_1 \ldots j_k, i}^\sigma - y_{j_1 \ldots j_k, i}^\sigma$. The matrix of the subsystem of (5.3) with $r \leq s \leq 2r-1$ has the form

$$(5.4) \qquad \begin{pmatrix} \dfrac{\partial P_\sigma^i}{\partial y_{p_1 \ldots p_r}^\nu} & \cdots & \dfrac{\partial P_\sigma^{ij_1 \ldots j_{r-1}}}{\partial y_{p_1 \ldots p_r}^\nu} \\ \dfrac{\partial P_\sigma^i}{\partial y_{p_1 \ldots p_{r+1}}^\nu} & \cdots \dfrac{\partial P_\sigma^{ij_1 \ldots j_{r-2}}}{\partial y_{p_1 \ldots p_{r+1}}^\nu} & 0 \\ \cdots & & \\ \dfrac{\partial P_\sigma^i}{\partial y_{p_1 \ldots p_{2r-1}}^\nu} & & 0 \end{pmatrix}$$

and is of maximal rank, by the regularity assumption. Hence this subsystem has the only solution

$$(5.5) \qquad \frac{\partial(y^\sigma \circ \delta)}{\partial x^i} - y_i^\sigma \circ \delta = 0, \ldots, \frac{\partial(y_{j_1 \ldots j_{r-1}}^\sigma \circ \delta)}{\partial x^i} - y_{j_1 \ldots j_{r-1} i}^\sigma \circ \delta = 0$$

which implies that

$$(5.6) \qquad \pi_{2r-1,r} \circ \delta = j^r(\pi_{2r-1,0} \circ \delta).$$

Substituting these equalities in the remaining equations (5.3) we obtain that $\gamma = \pi_{2r-1,0} \circ \delta$ satisfies the Euler-Lagrange equations of $\lambda$, i.e., is an extremal.

Now if $\gamma$ is given and the assumptions of Theorem 5 hold, $\gamma$ restricted to a neighborhood of each point of its domain of definition must be an extremal; hence $\gamma$ is an extremal as required.

A sufficient condition of regularity has been given by Shadwick [20]. To formulate this condition we adopt the convention that the underlining of some indices means symmetrization in these indices.

**Theorem 6.** *Let $\lambda \in \Omega_X^n(j^rY)$ be a lagrangian. Suppose that to each point $y \in Y$ there exists a fiber chart $(V,\psi)$, $\psi = (x^i, y^\sigma)$, on $Y$, for which $\lambda$ has an expression $\lambda = L\omega_0$, such that $y \in V$, and all the matrices*

$$(5.7) \quad \frac{1}{N(j_1\ldots j_{2r-s}\underline{p_{r+1}}\ldots\underline{p_s})N(\underline{p_1}\ldots\underline{p_r})} \frac{\partial^2 L}{\partial y^\sigma_{j_1\ldots j_{2r-s}\underline{p_{r+1}}\ldots\underline{p_s}} \partial y^\nu_{\underline{p_1}\ldots\underline{p_r}}}, \quad r \leq s \leq 2r-1,$$

*whose the columns (resp. rows) are labeled by the sequences $j_1 \leq \ldots \leq j_{2r-s}$ (resp. $p_1 \leq \ldots \leq p_s$), are of maximal rank at each point of $V_r$. Then $\lambda$ is regular.*

**Proof.** Theorem 6 follows from the fact that the matrices (5.7) are precisely the diagonal submatrices in (5.4).

The regularity condition of Theorem 6 is independent of the choice of a fiber chart $(V,\psi)$.

Theorem 5 shows that if a lagrangian is regular, its extremals are characterized, locally, by means of Hamilton extremals of the local Poincare-Cartan equivalents (*local Hamilton extremals*); locally, there is a bijective correspondence between extremals of the lagrangian and some *classes* of local Hamilton extremals.

An analogous global result is obtained by means of Lepagean equivalents of Theorem 3 provided the assumptions of Theorem 6 hold.

**Theorem 7.** *Let $\lambda \in \Omega_X^n(j^rY)$ be a lagrangian satisfying the hypothesis of Theorem 6. Let $\rho$ be a Lepagean equivalent of $\lambda$ such that the assumptions of Theorem 3 hold. Then for each Hamilton extremal $\delta$ of $\rho$ the section $\gamma = \pi_{2r-1,0} \circ \delta$ of $\pi$ is an extremal of $\lambda$.*

**Proof.** This theorem is proved in full analogy with the proof of Theorem 5, in which the local Poincare-Cartan form $\Theta_{\lambda,V}$ is replaced by $\rho$.

**6. The Hamilton-de Donder equations.** Let $\lambda \in \Omega_X^n(j^rY)$ be a lagrangian, $j_x^{2r-1}\gamma \in j^{2r-1}Y$ a regular point of $\lambda$. There exists a fiber chart $(V,\psi)$, $\psi = (x^i, y^\sigma)$, on $Y$ such that $\gamma(x) \in V$, and the functions $x^i$, $y^\sigma$, $y^\sigma_{j_1}$, ..., $y^\sigma_{j_1\ldots j_{r-1}}$, $P^{j_1}_\sigma$, ..., $P^{j_1\ldots j_r}_\sigma$ (see (3.6)), where $j_1 \leq \ldots \leq j_r$, form a part of a chart $(W,\Psi)$ on $j^{2r-1}Y$ about $j_x^{2r-1}\gamma$. The local Poincare-Cartan equivalent $\Theta_{\lambda,V}$ is expressed for the chart $(W,\Psi)$ by (3.7), where $L$ is considered as the function of $x^i$, $y^\sigma$, $y^\sigma_{j_1}$, ..., $y^\sigma_{j_1\ldots j_{r-1}}$, $P^{j_1\ldots j_r}_\sigma$. Setting

$$(6.1) \quad H = L + \sum_{k=1}^{r} \sum_{j_1 \leq \ldots \leq j_k} N(j_1\ldots j_k) P^{j_1\ldots j_k}_\sigma y^\sigma_{j_1\ldots j_k}$$

we obtain the following expression for $\Theta_{\lambda,V}$:

$$(6.2) \quad \Theta_{\lambda,V} = H\omega_0 + \left(P^i_\sigma dy^\sigma + \sum_{k=1}^{r-1} \sum_{j_1 \leq \ldots \leq j_k} N(j_1\ldots j_k) P^{ij_1\ldots j_k}_\sigma dy^\sigma_{j_1\ldots j_k}\right) \wedge \omega_i.$$

The expression of the Hamilton form $H_{\lambda,V}$ of $\Theta_{\lambda,V}$ is then immediately obtained from the definition (4.1). We get

$$
\begin{aligned}
H_{\lambda,V} = -\bigg[&\left(\frac{\partial H}{\partial y^\sigma} + d_i P_\sigma^i\right) dy^\sigma \\
&+ \sum_{k=1}^{r-1} \sum_{j_1 \leq \ldots \leq j_k} \left(\frac{\partial H}{\partial y^\sigma_{j_1 \ldots j_k}} + N(j_1 \ldots j_k) d_i P_\sigma^{ij_1 \ldots j_k}\right) dy^\sigma_{j_1 \ldots j_k} \\
&+ \sum_{k=1}^{r} \sum_{j_1 \leq \ldots \leq j_k} \left(\frac{\partial H}{\partial P_\sigma^{j_1 \ldots j_k}} - N(j_1 \ldots j_k) y^\sigma_{(j_1 \ldots j_{k-1}, j_k)}\right) dP_\sigma^{j_1 \ldots j_k}\bigg] \wedge \omega_0.
\end{aligned}
\tag{6.3}
$$

**Theorem 8.** *Let $\lambda \in \Omega^n_X(j^r Y)$ be a regular lagrangian, $j_x^{2r-1}\gamma \in j^{2r-1}Y$ its regular point, $(V,\psi)$, $\psi = (x^i, y^\sigma)$, a fiber chart on $Y$ such that $\gamma(x) \in V$, and the functions $x^i, y^\sigma, y^\sigma_{j_1}, \ldots, y^\sigma_{j_2 \ldots j_{r-1}}, P_\sigma^{j_1}, \ldots, P_\sigma^{j_1 \ldots j_r}$, $j_1 \leq \ldots \leq j_r$, form a part of a chart $(W, \Psi)$ on $j^{2r-1}Y$ about $j_x^{2r-1}\gamma$. Let $U = \pi(V)$. Then a section $\delta$ of $\pi_{2r-1}$ over $U$ such that $\delta(U) \subset W$, is a local Hamilton extremal of $\lambda$ if and only if*

$$
\frac{\partial H}{\partial y^\sigma} \circ \delta + \frac{\partial (P_\sigma^i \circ \delta)}{\partial x^i} = 0,
$$

$$
\frac{\partial H}{\partial y^\sigma_{j_1 \ldots j_k}} \circ \delta + N(j_1 \ldots j_k) \frac{\partial (P_\sigma^{ij_1 \ldots j_k} \circ \delta)}{\partial x^i} = 0,
$$

$$
1 \leq k \leq r-1,
\tag{6.4}
$$

$$
\frac{\partial H}{\partial P_\sigma^{j_1 \ldots j_k}} \circ \delta - \frac{N(j_1 \ldots j_k)}{k} \left(\frac{\partial (y^\sigma_{j_2 \ldots j_k} \circ \delta)}{\partial x^{j_1}} + \ldots + \frac{\partial (y^\sigma_{j_1 \ldots j_{k-1}} \circ \delta)}{\partial x^{j_k}}\right) = 0,
$$

$$
1 \leq k \leq r.
$$

**Proof.** Theorem 8 follows from the chart expression (6.3) of $H_{\lambda,V}$ and the definition of the Hamilton extremal.

Equations (6.4) were derived by de Donder [4]. We have related these equations to a "fundamental form"- the local Poincare-Cartan form, and interpret them as the Euler-Lagrange equations of the "extended lagrangian" (4.6) (with $\rho$ replaced by $\Theta_{\lambda,V}$) and, in accordance with [20], where a special regularity condition has been applied, as equations which can be used to obtain all local extremals of the initial lagrangian. We call these equations the *Hamilton-de Donder equations.*

**7. Extremal fields and the Hamilton-Jacobi equation.** Let $\xi$ be a vector field on $j^{2r-1}Y$, $w$ a global section of $\pi_{2r-1,r-1}$. Set for each $j_x^{r-1}\gamma \in j^{r-1}Y$

$$
\xi_w(j_x^{r-1}\gamma) = T\pi_{2r-1,r-1} \cdot \xi(w(j_x^{r-1}\gamma)).
\tag{7.1}
$$

$\xi_w$ is a vector field on $j^{r-1}Y$.

**Lemma.** *Let $\lambda \in \Omega^n_X(j^r Y)$ be a regular lagrangian, satisfying the assumptions of Theorem 6, $\rho$ a Lepagean equivalent of $\lambda$ of the form (3.9). Let $\gamma$ be a section of $\pi$, and suppose that*

(7.2) $\qquad \pi_{2r-1,r} \circ w \circ j^{r-1}\gamma = j^r \gamma.$

*Then for each vector field $\xi$ on $j^{2r-1}Y$*

(7.3) $\qquad j^{r-1}\gamma^* w^* i_\xi d\rho = j^{r-1}\gamma^* i_{\xi_w} w^* d\rho.$

**Proof.** Let $x \in X$ be a point from the domain of definition of $\gamma$, $\zeta_1, \ldots, \zeta_n \in T_x X$ tangent vectors. By a direct calculation

(7.4)
$$(j^{r-1}\gamma^* w^* i_\varphi d\rho)(x)(\zeta_1,\ldots,\zeta_n) = (j^{r-1}\gamma^* i_{\varphi_w} w^* d\rho)(x)(\zeta_1,\ldots,\zeta_n) +$$
$$d\rho(w(j_x^{r-1}\gamma))(\zeta(w(j_x^{r-1}\gamma)), Tw \cdot Tj^{r-1}\gamma \cdot \zeta_1, \ldots, Tw \cdot Tj^{r-1}\gamma \cdot \zeta_n),$$

where

(7.5) $\qquad \zeta(w(j_x^{r-1}\gamma)) = \xi(w(j_x^{r-1}\gamma)) - Tw \cdot \xi_w(j_x^{r-1}\gamma)$

is a $\pi_{2r-1,r-1}$-vertical vector. We have to show that the second term in (7.4) vanishes. Let $(V,\psi), \psi = (x^i, y^\sigma)$, be a fiber chart on $Y$ such that $\gamma(x) \in V$, and write $\rho$ (3.9) in the form (3.2). Then

(7.6)
$$d\rho = dL \wedge \omega_0 + \sum_{k=0}^{r-1} \sum_{j_1 \leq \ldots \leq j_k} df_\sigma^{i,j_1\ldots j_k} \wedge \omega^\sigma_{j_1\ldots j_k} \wedge \omega_i$$
$$\sum_{k=0}^{r-1} \sum_{j_1 \leq \ldots \leq j_k} f_\sigma^{i,j_1\ldots j_k} \omega^\sigma_{j_1\ldots j_k i} \wedge \omega_0,$$

and contracting this form at the point $w(j_x^{r-1}\gamma) \in j^{2r-1}Y$ by $\zeta(w(j_x^{r-1}\gamma))$ (7.5) we get

(7.7) $\qquad i_{\zeta(w(j_x^{r-1}\gamma))} d\rho = \sum_{k=0}^{r-1} \sum_{j_1 \leq \ldots \leq j_k} i_{\zeta(w(j_x^{r-1}\gamma))} df_\sigma^{i,j_1\ldots j_k} \omega^\sigma_{j_1\ldots j_k}(w(j_x^{r-1}\gamma)) \wedge \omega_i(x).$

Using (7.2) we obtain, for any $\zeta_i$, $1 \leq i \leq n$, and $0 \leq k \leq r-1$,

(7.8)
$$\omega^\sigma_{j_1\ldots j_k}(w(j_x^{r-1}\gamma))(Tw \cdot Tj^{r-1}\gamma \cdot \zeta_i)$$
$$= \omega^\sigma_{j_1\ldots j_k}((\pi_{2r-1,r} \circ w)(j_x^{r-1}\gamma))(T(\pi_{2r-1,r} \circ w) \circ Tj^{r-1}\gamma \cdot \zeta_i)$$
$$= j^r\gamma^* \omega^\sigma_{j_1\ldots j_k}(x)(\zeta_i) = 0.$$

Hence the second term in (7.4) vanishes, as required.

**Theorem 9.** *Let $\lambda \in \Omega^n_X(j^r Y)$ be a regular lagrangian, satisfying the assumptions of Theorem 6, $\rho$ a Lepagean equivalent of $\lambda$ of the form (3.9). Suppose that we have a global section $w$ of $\pi_{2r-1,r-1}$ such that*

(7.9)     $w^*d\rho = 0.$

*Then each section $\gamma$ of $\pi$ such that* (7.2) *holds, is an extremal of $\lambda$.*

**Proof.** Under our assumption, $j^{r-1}\gamma^*w^*i_\xi d\rho = 0$ for all vector fields $\xi$ on $j^{2r-1}Y$, by (7.3), and $\delta = w \circ j^{r-1}\gamma$ is a Hamilton extremal; now we apply Theorem 7.

A section $w$ of $\pi_{2r-1,r-1}$, satisfying (7.9), generalizes the notion of a *geodesic field*, or an *extremal field* from the first order theory [7], [8].

A section $w$ is a geodesic field if and only if, locally, there exists an $(n-1)$-form $S$ such that

(7.10)    $w^*\rho = dS.$

This is a modification of the *Hamilton-Jacobi equation* obtained by Shadwick [20] who used the local Poincare-Cartan equivalent $\Theta_\lambda$ in place of the global Lepagean equivalent $\rho$. We note that equation (7.10) splits in an equivalent system of equations

(7.11)    $h(w^*\rho) = h(dS), p_1(w^*\rho) = p_1(dS), p_2(dS) = 0, \ldots, p_n(dS) = 0.$

Let $w$ be a geodesic field, and suppose for simplicity that there exists an $(n-1)$-form $S$, defined on $j^{r-1}Y$, such that (7.10) holds. Let $\Omega \subset X$ be a piece, $\gamma$ a section of $\pi$ over $\Omega$. We set

(7.12)    $W_\Omega(\gamma) = \int_\Omega j^{r-1}\gamma^*w^*\rho.$

$W_\Omega$ is a generalization of the *Hilbert independent integral* [7]. If $\gamma_0$ is a section of $\pi$ over $\Omega$, such that

(7.13)    $\pi_{2r-1,r} \circ w \circ j^{r-1}\gamma_0 = j^r\gamma_0,$

and $\gamma$ is an arbitrary section of $\pi$ over $\Omega$ such that

(7.14)    $j^{r-1}\gamma|_{\partial\Omega} = j^{r-1}\gamma_0|_{\partial\Omega},$

we have

(7.15)    $\lambda_\Omega(\gamma_0) = \int_\Omega (w \circ j^{r-1}\gamma_0)^*\rho = W_\Omega(\gamma_0),$

(7.16)    $W_\Omega(\gamma) = \int_{\partial\Omega} j^{r-1}\gamma^*S = \int_{\partial\Omega} j^{r-1}\gamma_0^*S = W_\Omega(\gamma_0),$

and

(7.17)    $\lambda_\Omega(\gamma) - \lambda_\Omega(\gamma_0) = \lambda_\Omega(\gamma) - W_\Omega(\gamma) - \lambda_\Omega(\gamma_0) + W_\Omega(\gamma) = \int_\Omega j^r\gamma^*E_{\rho,w},$

where

(7.18)    $E_{\rho,w} = \lambda - \pi_{r,r-1}^*w^*\rho.$

This form generalizes the Weierstrass $E$-function, and may be called the *Weierstrass form* associated with $\rho$ and $w$. The horizontal part of the Weierstrass form can be used for the study of local extrema of the function $\gamma \to \lambda_\Omega(\gamma)$ on sections $\gamma$ defined by (7.14), relative to the $C^{r-1}$-topology (referred to here as the *strong local extrema*). Notice that by Theorem 9, $\gamma_0$ is necessarily an extremal.

**Theorem 10.** *Let $\lambda \in \Omega^n_X(j^rY)$ be a lagrangian, satisfying the assumptions of Theorem 6, $\rho$ a Lepagean equivalent (3.9) of $\lambda$, $w$ a geodesic field, $\Omega \subset X$ a piece, and $\gamma_0$ a section of $\pi$ over $\Omega$ such that (7.13) holds. Suppose that there exists an open neighborhood $Z$ of $j^{r-1}\gamma_0(\Omega)$ in $j^{r-1}Y$ such that the form $h(E_{\rho,w})$ is non-negative on $\pi^{-1}_{r,r-1}(Z) \subset j^rY$ relative to the orientation of $X$. Then $\gamma_0$ is a strong local minimum for $\lambda_\Omega$.*

**Proof.** This follows from (7.17) rewritten in the form

$$(7.19) \quad \lambda_\Omega(\gamma) = \lambda_\Omega(\gamma_0) + \int_\Omega j^r\gamma^* h(E_{\rho,w}),$$

where $\gamma$ is any section of $\pi$ over $\Omega$ satisfying (7.14), such that $j^{r-1}\gamma(\Omega) \subset Z$.

Let $\lambda \in \Omega^n_X(j^rY)$ be a lagrangian, $j^r_x\gamma \in j^rY$ a point, $VT_{j^r_x\gamma}j^rY$ the vector space of $\pi_{r,r-1}$-vertical tangent vectors to $j^rY$ at $j^r_x\gamma$. Let $(V,\psi)$, $\psi = (x^i, y^\sigma)$, be a fiber chart on $Y$, positive with respect to the orientation of $X$, such that $\gamma(x) \in V$. For any vector $\zeta \in VT_{j^r_x\gamma}j^rY$ we set

$$(7.20) \quad H_\lambda(\zeta) = \left( \sum_{j_1 \leq \ldots \leq j_r} \sum_{q_1 \leq \ldots \leq q_r} \frac{\partial^2 L}{\partial y^\sigma_{j_1\ldots j_r} \partial y^\nu_{q_1\ldots q_r}} \cdot \zeta^\sigma_{j_1\ldots j_r} \zeta^\nu_{q_1\ldots q_r} \right) \omega_0,$$

where

$$(7.21) \quad \lambda = L\varphi_0, \quad \zeta = \sum_{j_1 \leq \ldots \leq j_r} \zeta^\sigma_{j_1\ldots j_r} \frac{\partial L}{\partial y^\sigma_{j_1\ldots j_r}}$$

with respect to $(V,\psi)$, and the right side is considered at the point $j^r_x\gamma$. $H_\lambda(\zeta)$ is a well-defined quadratic form on $VT_{j^r_x\gamma}j^rY$ with values in $\wedge^n T^*_xX$. We say that $\lambda$ is *positive definite* at $j^r_x\gamma$ if $H_\lambda(\zeta) > 0$ (relative to the orientation of $X$) for each $\zeta \neq 0$. Obviously $\lambda$ is positive definite at $j^r_x\gamma$ if and only if the matrix

$$(7.22) \quad \left( \frac{\partial^2 L}{\partial y^\sigma_{j_1\ldots j_r} \partial y^\nu_{q_1\ldots q_r}} \right), \quad j_1 \leq \ldots \leq j_r, q_1 \leq \ldots \leq q_r,$$

whose rows (resp. columns) are labeled by $\binom{\sigma}{j_1\ldots j_r}$ (resp. $\binom{\nu}{q_1\ldots q_r}$) is positive definite at $j^r_x\gamma$. $\lambda$ is said to be *positive definite* if it is positive definite at each point $j^r_x\gamma \in j^rY$.

Clearly, the notion of a positive definite lagrangian may be equivalently introduced with the help of a volume element on $X$; then (7.20) is replaced by a real valued quadratic form.

In the following theorem we consider local extrema of the function $\gamma \to \lambda_\Omega(\gamma)$

on section $\lambda$ of $\pi$, satisfying fixed boundary conditions of the form (7.14), relative to the $C^r$-topology (referred to as the *weak local extrema*).

**Theorem 11.** *Let* $\lambda \in \Omega_X^n(j^r Y)$ *be a positive lagrangian,* $\rho$ *a Lepagean equivalent* (3.9) *of* $\lambda$, $w$ *a global sections of* $\pi_{2r-1,r-1}$, $\Omega \subset X$ *a piece, and* $\gamma_0$ *a section of* $\pi$ *over* $\Omega$. *Suppose that the following conditions hold:*

(1) $\lambda$ *satisfies the assumptions of Theorem 6, and is positive definite at each point of* $\pi_{2r-1,r} \circ w(j^{r-1}Y)$,

(2) *there exists an* $(n-1)$-*form* $S$ *on* $j^{r-1}Y$ *such that* $w^*\rho = dS$,

(3) $\gamma_0$ *satisfies* $\pi_{2r-1,r} \circ w \circ j^{r-1}\gamma_0 = j^r \gamma_0$.

*Then* $\gamma_0$ *is a weak local minimum for* $\lambda_\Omega$.

**Proof.** Let $U \subset \mathbf{R}^n$, $V \subset \mathbf{R}^m$ be open sets, $f: U \times V \to \mathbf{R}$ a smooth function, $w_0: U \to V$ a smooth mapping, $w_0 = (w_0^\sigma)$. Suppose that the following conditions hold for each $x \in U$: (1) $f(x, w_0(x)) = 0$, (2) $Df(x, w_0(x)) = 0$, (3) the matrix

$$(7.23) \quad \left( \frac{\partial^2 f}{\partial y^\sigma \partial y^\nu} \right)_{(x, w_0(x))},$$

where $y^1, \ldots, y^m$ are the standard coordinates on $V$, is positive definite. Since all principal minors of this matrix are positive definite, the continuity arguments show that this matrix is positive definite on a neighborhood of each point $(x, w_0(x))$. We choose the product of open rectangles for this neighborhood, and denote by $W$ the union of these open rectangles; the matrix (7.23) is positive definite on $W$. Let $(x, y) \in W$. By definition of $W$, the point $(x, y)$ and $(x, w_0(x))$ can be joint by a segment. Using the Taylor's formula we get

$$(7.24) \quad \begin{aligned} f(x, y) &= f(x, w_0(x)) + \left( \frac{\partial f}{\partial y^\sigma} \right)_{(x, w_0(x))} (y^\sigma - w_0^\sigma(x)) \\ &+ \frac{1}{2} \left( \frac{\partial^2 f}{\partial y^\sigma \partial y^\nu} \right)_{(x, \zeta)} (y^\sigma - w_0^\sigma(x))(y^\nu - w_0^\nu(x)), \end{aligned}$$

where $\zeta$ is a point of the segment. Since $(x, \zeta) \in W$, the matrix

$$(7.25) \quad \left( \frac{\partial^2 f}{\partial y^\sigma \partial y^\nu} \right)_{(x, \zeta)}$$

is positive definite, and conditions (1), (2) imply

$$(7.26) \quad f(x, y) = \frac{1}{2} \left( \frac{\partial^2 f}{\partial y^\sigma \partial y^\nu} \right)_{(x, \zeta)} (y^\sigma - w_0^\sigma(x))(y^\nu - w_0^\nu(x)) \geq 0.$$

Consequently, each of the points $(x, w_0(x)) \in U \times V$ is a local minimum for $f$.

Let $(V, \psi)$, $\psi = (x^i, y^\sigma)$, be a fiber chart on $Y$, positive with respect to the orientation of $X$. In this chart, $h(E_{\rho, w}) = e_w(L)\omega_0$, where

$$e_w(L) = L - L_0 \pi_{2r-1,r} w \pi_{r,r-1}$$
(7.27)
$$- \sum_{j_1 \leq \ldots \leq j_r} \left( \frac{\partial L}{\partial y^\sigma_{j_1 \ldots j_r}} \circ \pi_{2r-1,r} w \pi_{r,r-1} \right) \left( y^\sigma_{j_1 \ldots j_r} - y^\sigma_{j_1 \ldots j_r} \circ \pi_{2r-1,r} w \pi_{r,r-1} \right),$$

and we apply the above remarks to the function $e_w(L): V_r \to \mathbf{R}$. We write $V_r = V_{r-1} \times Z$, and represent the section $\pi_{2r-1,r} \circ w$ of $\pi_{r,r-1}$ by the components $w_0 = (w^\sigma_{j_1 \ldots j_r}): V_{r-1} \to Z$. It is directly verified that for each $j^r_x \gamma \in w(V_{r-1}) = V_{r-1} \times w_0(V_{r-1})$

(7.28) $\quad e_w(L)(j^r_x \gamma) = 0, \quad \left( \frac{\partial e_w(L)}{\partial x^i} \right)_{j^r_x \gamma} = 0, \quad \left( \frac{\partial e_w(L)}{\partial y^\nu_{q_1 \ldots q_k}} \right)_{j^r_x \gamma} = 0, \quad 0 \leq k \leq r,$

and

(7.29) $\quad \left( \frac{\partial^2 e_w(L)}{\partial y^\rho_{p_1 \ldots p_r} \partial y^\nu_{q_1 \ldots q_r}} \right)_{j^r_x \gamma} = \left( \frac{\partial^2 L}{\partial y^\rho_{p_1 \ldots p_r} \partial y^\nu_{q_1 \ldots q_r}} \right)_{j^r_x \gamma}.$

In particular, the matrix (7.29) is positive definite. Replacing in the above discussion $U$ (resp. $V$, resp. $f$) by $V_{r-1}$ (resp. $Z$, resp. $e_w(L)$) we obtain that $e_w(L) \geq 0$ on a neighborhood of $\pi_{2r-1,r}(w(V_{r-1})) \subset V_r$; consequently,

(7.30) $\quad h(E_{\rho,w}) \geq 0,$

(relative to the orientation of $X$) on a neighborhood $W$ of $\pi_{2r-1,r}(w(j^{r-1}Y) \subset j^r Y$.

Let us consider the section $\gamma_0$ of $\pi$. Conditions (1) - (3) of Theorem 11 guarantee that $\gamma_0$ is an extremal of $\lambda_\Omega$ (Theorem 9). Since $j^r \gamma_0(\Omega) = \pi_{2r-1,r}(w(j^{r-1}\gamma_0(\Omega))) \subset \pi_{2r-1,r}(w(j^{r-1}Y)) \subset W$, our assertion follows from (7.17).

J.E. Purkyne University, Faculty of Science
Department of Mathematics
Janackovo nam. 2a, 662 95 Brno, Czechoslovakia